\documentclass{elsart3p}

 \usepackage{graphicx}
\usepackage{amssymb}

\begin{document}

\begin{frontmatter}



\title{Electromagnetic Probes in PHENIX}


\author{G. David for the PHENIX collaboration\thanksref{authList}}
\thanks[authList]{For the full list of PHENIX authors and 
 acknowledgements, see for instance~\cite{ppg042}.}
\address{Brookhaven National Laboratory, Upton, NY 11973}
\thanks[ead]{{\it Email address:} david@bnl.gov}

\begin{abstract}
% Text of abstract
Electromagnetic probes are arguably the most universal tools to study
the different physics processes in high energy hadron and heavy ion collisions.
In this paper we summarize recent measurements of real and virtual
direct photons at central rapidity by the PHENIX experiment at RHIC 
in p+p, d+Au and Au+Au collisions.  
We also discuss the impact of the results and the constraints they put on
theoretical models.  At the end we report on the immediate as well as
on the mid-term future of photon measurements in PHENIX.

\end{abstract}

\end{frontmatter}
% main text
\section{Introduction}
\label{sec_intro}

Electromagnetic probes are of crucial importance in relativistic heavy
ion physics.  They are real and virtual direct photons (where 
{\it  direct} means that they are not produced in the decay of some
final state hadron), and they are penetrating
probes produced in virtually all subprocesses of relativistic heavy
ion collisions.  Equally important, once produced they escape
unaltered even from a very dense medium because of their weak coupling
($\alpha_e<<\alpha_s$).  
However, this very same property that makes
them excellent ``historians'' of the entire collision makes both
precise measurements and drawing of definitive conclusions
challenging, since one has to disentangle contributions from different
processes often with comparable rates in the same $p_T$ region.  

In fact, for the experimenter difficulties start already by finding 
the {\it direct} photons themselves in the large background
of photons from (final state) hadron decays like
$\pi^0, \eta$, {\it etc.}.  This can be especially hard in the low 
$p_T$ region; at higher $p_T$ it is somewhat easier due to the 
large suppression of
hadrons observed in heavy ion collisions (Sec.~\ref{sec_pi0}).
Interpretation of the results is challenging because the net signal
is a convolution of photon spectra from a large variety 
of possible physics processes as different as initial hard scattering,
a pure pQCD process, jet fragmentation, Bremsstrahlung, jet-photon 
conversion in the medium,
and thermal radiation from the quark-gluon plasma (QGP) 
and/or a hot hadron gas.  

Despite all these difficulties, photon physics made major inroads in
the first few years of RHIC, both experimentally and theoretically -
although not always in the areas originally anticipated.  
On one hand NLO pQCD
calculations of photoproduction at central rapidities have been
confirmed both in p+p and d+Au (Sec.~\ref{sec_ppdAu}), and we also
have a better understanding of the ratio of photons from initial hard
scattering and fragmentation in p+p.  
On the other hand one of the
major pre-RHIC expectations was to relatively quickly establish the
initial temperature $T_i$ after the collision from thermal radiation; 
data currently available (Sec.~\ref{sec_auau}) are well described 
(within a factor of 2) by various models in which $T_i$ ranges 
from 300 to 660 MeV - but this uncertainty is largely due to the 
unexpectedly fast thermalization (Sec~\ref{sec_intconv}).  
In Au+Au
collisions the non-suppression of photons (Sec.~\ref{sec_auau}) became
an independent validation of pQCD and of the concept of $T_{AA}$ 
scaling when establishing ``jet quenching'' in hadrons 
(Sec.~\ref{sec_pi0}).  Also in Au+Au at medium $p_T$ there are
hints of medium-induced photons, possibly jet-photon
conversion on thermal quarks (Sec.~\ref{sec_auau}) and in the region
where thermal radiation is expected to dominate preliminary results
from a new analysis technique (applied for the first time in a heavy 
ion experiment) finds a possible signal (Sec.~\ref{sec_intconv}).  
One way to disentangle contributions from different processes like
fragmentation and jet-photon conversion is to study the azimuthal
asymmetries of photons as a function of $p_T$, which also can put
constraints of the jet energy loss mechanisms.  The first data on 
photon asymmetries are not conclusive yet (Sec.~\ref{sec_photonv2})
but new results with much higher statistics are expected to be
published soon.  Finally, imminent and future upgrades of PHENIX will
enhance the scope of our photon physics even further, among others
with a competitive measurement of vector mesons at central rapidities 
and direct photons in the low-$x$ region (Sec.~\ref{sec_outlook}).

In this paper we concentrate on real photons and dielectron pairs from
internal conversion of photons (other results with electromagnetic
probes are described in~\cite{dion,bickley,rak} in these Proceedings).
As for the detector,
the primary tools to detect photons in PHENIX are the
lead-scintillator (PbSc) and the lead-glass (PbGl) calorimeters,
described in detail in~\cite{nim2003}; experimental techniques are
briefly described in~\cite{tadaaki,dmitri}.  
Note that the two detectors are fundamentally different, as are their
associated systematic errors, their data are analyzed separately,
therefore, comparing their results is a strong, built-in consistency
check in PHENIX.  Electron pairs from virtual
photons are identified by the tracking system, the Ring Imaging
Cherenkov Detector (RICH) and the calorimeters~\cite{nim2003}.

\section{The ``background'': neutral mesons}
\label{sec_pi0}

\begin{figure}
\begin{center}
\includegraphics*[width=8cm]{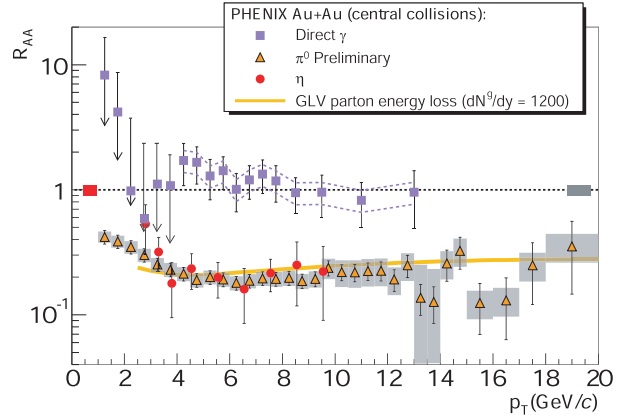}
\end{center}
\caption{Nuclear modification factor $R_{AA}$ in central Au+Au
  collisions ($\sqrt{s{NN}}=200$GeV) for $\pi^0$, $\eta$ and direct
  photons, compared to a GLV calculation with $dN^g/dy=1200$ gluon
  density.  The mesons - independent of their mass - are suppressed by
  a factor of 5 with respect to their yield in p+p scaled with the
  nuclear thickness $T_{AB}$, whereas the direct photons are not
  suppressed.   }
\label{fig:raa_AuAu200}
\end{figure}

Before discussing {\it direct} photons we should review briefly
what is known about neutral meson ($\pi^0$, $\eta$) production in
different systems colliding at RHIC energies.  In $\sqrt{s}=200$GeV
p+p collisions the measured $\pi^0$ cross section~\cite{pppi0}
is well described by the KKP set of fragmentation functions~\cite{KKP}
up to $p_T=13$GeV/c (recent preliminary results from Run-5 extend this
measurement and confirm the agreement with pQCD up to $p_T=18$GeV/c).
In order to characterize what is different in heavy ion collisions we
introduce the {\it nuclear modification factor} $R_{AA}$ 

\begin{equation}
R_{AA}(p_T)\;=\;\frac{dN_{AuAu}^{}/dp_{T}}{T_{AA}\cdot
d\sigma_{pp}/dp_{T}}
\end{equation}

\noindent
where $T_{AA}$ is the nuclear overlap function integrated over the
relevant impact parameter range.  The behavior of $R_{AA}$ at high
$p_T$ (in the pQCD regime), in specific, its deviation from unity 
is a very important, albeit somewhat ambigous indicator of the 
appearence of new, non-perturbative processes in nuclei.
In d+Au collisions and at least at central rapidities $R_{AA}$ 
for $\pi^0$ is consistent with unity~\cite{ppg028}.  The situation is
dramatically different in Au+Au collisions~\cite{ppg003,ppg014}:
$\pi^0$ production at midrapidity 
is suppressed by a factor of 5 with respect to $T_{AA}$-scaled pQCD 
expectations, and as seen of Fig.~\ref{fig:raa_AuAu200} the suppression 
is constant up to the highest $p_T$ where identified hadrons have 
been masured so far at RHIC.  Not only does this ``jet quenching''
constitute one of the earliest and most important discoveries at RHIC
but it also makes the direct photon measurement somewhat easier at
high $p_T$ since the S/B is significantly improved.  Furthermore,
PHENIX has also shown that the suppression pattern is the same for
$\eta$ mesons~\cite{ppg051}, the second most important background for
direct photons (Fig.~\ref{fig:raa_AuAu200}).  Finally, PHENIX 
also masured $R_{AA}$ at $\sqrt{s_{NN}}=200$GeV in Cu+Cu collisions, a
system much smaller than Au+Au.  When mid-peripheral Au+Au
collisions were compared to central Cu+Cu collisions (characterized by
the same number of participants $N_{part}$, {\it i.e.} by the same
overlap volume albeit by very different geometry), the suppression 
pattern was the same.  We should point out that this is true for 
the {\it average}, $\phi$-integrated $R_{AA}$ only like the one shown
on Fig.~\ref{fig:raa_AuAu200}.  If one studies $R_{AA}$ as a function
of the angle $\Delta\phi$ with respect to the event-by-event reaction 
plane~\cite{winter2005} a significant angular anisotropy is observed (and
attributed to the difference in average pathlength in the medium for
quarks in or out of the reaction plane).  So we fully expect that a
comparison of the differential $R_{AA}(\Delta\phi,p_T)$ in the
symmetric central Cu+Cu and the ($N_{part}$-equivalent) eccentric
Au+Au systems will reveal differences and shed some more light on the
nature of jet energy loss.  

Hadron suppression in Au+Au
and lack thereof in d+Au collisions together formed compelling
evidence that a very high density matter has been formed at RHIC that
strongly modifies jets created in the initial hard collisions and that
- at least at central rapidities - these modifications are not due to
a change of the (initial) parton distribution functions in a relativistic
nucleus.  However, from a purist's point of view neither the hadron
suppression nor conclusions drawn from it are on solid footing until
the validity of the concept ``scaling p+p cross-sections by $T_{AA}$''
is proven with a penetrating probe.  This happened by measuring 
$R_{AA}$ for direct photons as also shown on Fig.~\ref{fig:raa_AuAu200}:
at and above $\sim5$GeV/c the direct photon $R_{AA}$ is consistent
with unity.

\section{Direct photons in p+p and d+Au collisions}
\label{sec_ppdAu}

In p+p collisions to leading order in $\alpha_s$ direct photon
production is dominated by gluon Compton-scattering 
($q+g \rightarrow q+\gamma$), and since RHIC is the world's only
polarized p+p collider it is a unique place to study (polarized) gluon
structure functions.  However, as pointed out in~\cite{dmitri} photons
emitted in parton fragmentation are also an important source of direct
photons, particularly at low $p_T$.  In fact, calculations based upon
CTEQ6 parton distribution functions and the GRV fragmentation function
set predict that at $\sqrt{s_{NN}}=200$GeV and at transverse momenta 
below 3GeV/c more than half of photons will come from fragmentation.
Note that this is the very same $p_T$ region where in heavy ion
collisions the thermal photon signal from the QGP, even if not
dominant, might be sufficiently strong to be visible 
(Fig.~\ref{fig:phot-phenix-theo} from~\cite{turbide2004}).  Therefore,
measurement and theoretical understanding of photon production in p+p
is crucial both in its own right and as a baseline to interpret Au+Au
data. 

\begin{figure}
\begin{center}
\includegraphics*[width=8cm]{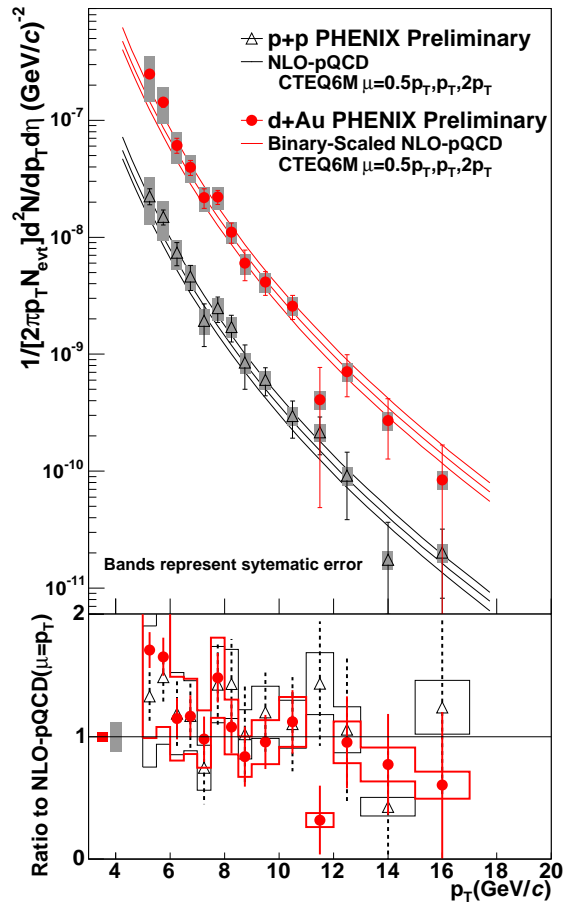}
\end{center}
\caption{Inclusive direct photon yield measured in $\sqrt{s_{NN}}$ 
  p+p (open triangles) and  d+Au (solid circles) collisions and
  compared to NLO pQCD calculations.     }
\label{fig:d_Au-pp-comp}
\end{figure}

The PHENIX results from Run-3 on the inclusive direct photon yield 
(prompt and fragmentation) 
for $\sqrt{s_{NN}}=200$GeV p+p and d+Au collisions is
shown on Fig.~\ref{fig:d_Au-pp-comp} along with NLO pQCD calculations
performed with three different renormalization scales~\cite{vogelsang}.
At the bottom the data/theory ratios are plotted: for both systems the
agreement is quite good, although not as spectacular as for 
pions in p+p~\cite{ppg024}.

Low multiplicities in p+p collisions allow PHENIX to tag isolated 
photons by requiring that the energy deposited in a cone around the
photon is small compared to the photon energy
itself~\cite{dmitri}.  Those isolated
photons come mostly from primordial hard scattering (as opposed to
fragmentation).  The ratio of isolated/inclusive direct photons has
been measured as well as calculated and it is a sensitive test of the
photon fragmentation functions~\cite{dmitri} although one has to
be careful to reproduce the acceptance bias of the experiment in the
theoretical calculation.   
At $p_T=7$GeV/c and above, {\it i.e.} in the region
where contribution from fragmentation is small, the measurement agrees 
with the calculation quite well; at small $p_T$ improved experimental
techniques are needed to make meaningful comparisons.

\section{Direct photons in Au+Au collisions}
\label{sec_auau}

Based upon the $\sqrt{s_{NN}}=200$GeV Au+Au data from Run-2 
PHENIX already published~\cite{ppg042} centrality dependent 
invariant photon yields
and photon excess double ratios up to $p_T=14$GeV/c and found 
- within sizeable systematic errors - that at
high transverse momenta ($p_T>5-6$GeV/c the data agree well with NLO
pQCD predictions scaled by $T_{AA}$.  (The widely used photon excess
ratio is the ratio $N^{inc}/N^{had}$ of the inclusive photons to
photons from hadron decays, with values above 1 indicating the
presence of direct photons.  For technical reasons it is often plotted
as the ''excess double ratio'' 
$\frac{N^{inc}/N^{\pi^0}}{N^{had}/N^{\pi^0}_{fit}}$ because it
eliminates some of the systematic errors.)

\begin{figure}
\begin{center}
\includegraphics*[width=8cm]{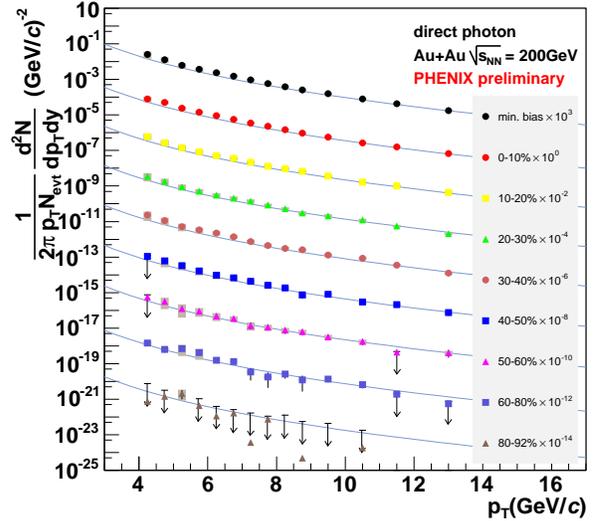}
\end{center}
\caption{Direct photon inveriant yields in $\sqrt{s_{NN}}=200$GeV
 Au+Au collisions, all centrality classes and minimum bias, compared
 to NLO pQCD calculations scaled by $T_{AA}$ (see also~\cite{tadaaki}).
     }
\label{fig:cdirphoton}
\end{figure}

The much larger
dataset in Run-4 allowed PHENIX to extend the $p_T$ range up to 18 GeV/c
and to reduce the errors at medium $p_T$ where several new photon
production mechanisms (beyond pQCD) have been suggested recently
including photons from jet-plasma interactions.  The preliminary
results from Run-4 on Fig.~\ref{fig:cdirphoton} show combined results
from the PHENIX PbSc and PbGl up to 14GeV/c (where the two analyses
are completely consistent, see Sec.~\ref{sec_intro}) and the results
are compared to $T_{AA}$-scaled NLO pQCD calculations; the overall
agreement is very good.

\begin{figure}
\begin{center}
\includegraphics*[width=8cm]{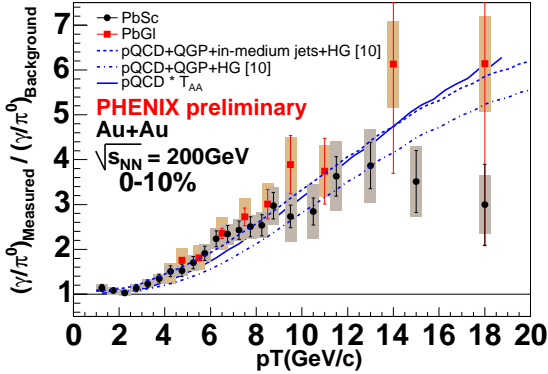}
\end{center}
\caption{Direct photon excess (double) ratio up to 
  $p_T=18$GeV/c measured in
  the PHENIX PbSc (solid circles) and PbGl (solid squares)
  calorimeters and compared to three theoretical 
  calculations~\cite{denterria}, where for the plasma 
  a formation time of
  $\tau_0=0.15$fm/c and an initial temperature of 
  $T_i=590$MeV is assumed   (see  also~\cite{tadaaki}).
     }
\label{fig:cdoubleratio_comp}
\end{figure}

A more detailed look to the most central collisions is given on
Fig.~\ref{fig:cdoubleratio_comp} where results for the two
calorimeters are plotted separately and up to 18GeV/c along with
$T_{AA}$-scaled NLO pQCD alone, with contributions from the QGP and
hadron gas and with in-medium jet modifications.  Note that in this
particular model a very short thermalization time and high initial
temperature are assumed ($\tau_0=0.15$fm/c and $T_i=590$MeV,
respectively); as shown later on Fig.~\ref{fig:phot-phenix-theo} they
are not the only possible choice to describe the PHENIX data.
Also, while strictly speaking the PbSc and PbGl results agree within
errors even above $p_T=14$GeV/c, there is an apparent ``deficit'' in
the PbSc with respect to both the PbGl and pQCD; the ongoing analysis
of the full Run-4 dataset should clarify this situation very soon.
The errors at medium $p_T$ (5-10GeV/c) are already smaller than in the
published Run-2 data~\cite{ppg042} but the trend that almost all
points are above $T_{AA}$-scaled pQCD survives - a tantalizing hint of
additional photon sources like jet-photon conversion, which can be
further explored investigating azimuthal asymmetries in the photon
distribution (Sec.~\ref{sec_photonv2}).  
We discuss the low $p_T$ region (1-5GeV/c) in
the context of the internal conversion measurement 
(Sec.~\ref{sec_intconv}).

\section{Internal conversion photons in Au+Au collisions}
\label{sec_intconv}

\begin{figure}
\begin{center}
\includegraphics*[width=8cm]{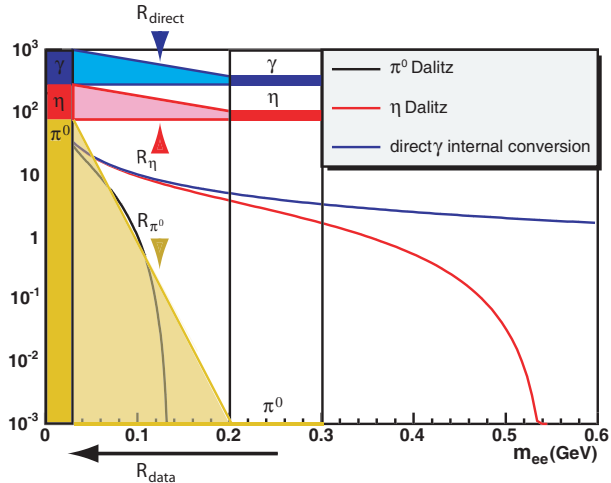}
\end{center}
\caption{Illustration of the direct photon measurement {\it via}
     low-mass dileptons
     }
\label{fig:akiba}
\end{figure}

\begin{figure}
\begin{center}
\includegraphics*[width=8cm]{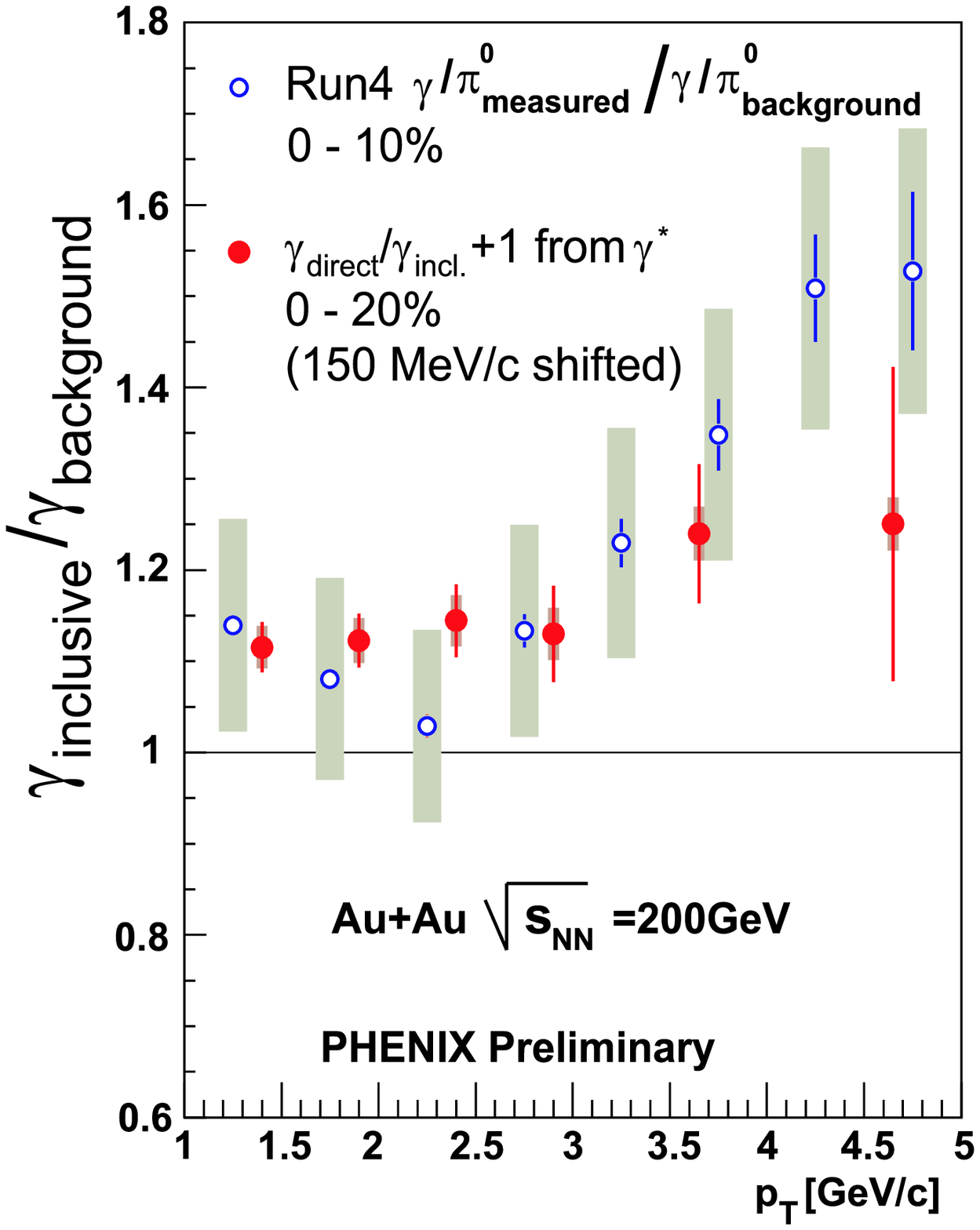}
\end{center}
\caption{Nuclear modification factor $R_{AA}$ in central Au+Au
  collisions ($\sqrt{s{NN}}=200$GeV) for $\pi^0$, $\eta$ and direct
  photons, compared to a GLV calculation with $dN^g/dy=1200$ gluon
  density.  The mesons - independent of their mass - are suppressed by
  a factor of 5 with respect to their yield in p+p scaled with the
  nuclear thickness $T_{AB}$, whereas the direct photons are not
  suppressed.   }
\label{fig:photon_excess_AuAu}
\end{figure}

A very promising approach to measure low-$p_T$ direct photons
is to utilize low-mass electron pairs from ``internal conversions'',
a technique first applied in heavy-ion collisions recently by
PHENIX~\cite{bathe2005}.  The basic idea is that any process
producing a real photon can also produce a virtual one of very low
mass~\cite{KrollWada}, subsequently decaying into an $e^+e^-$ pair.
This direct photon signal competes, of course, with dielectrons
from Dalitz decays of $\pi^0$, $\eta$, {\it etc}.
The rate and mass distributions of dielectrons are described both
for the low-mass direct photons and the Dalitz decays by the
Kroll-Wada formula~\cite{KrollWada},
  \begin{eqnarray}
    \lefteqn{\frac{1}{N_\gamma} \frac{dN_{ee}}{dm_{ee}}=}  \\ 
      & & \frac{2\alpha}{3\pi}
    \sqrt{1 - \frac{4m_e^2}{m_{ee}^2}} (1 + \frac{2m_e^2}{m_{ee}^2}) 
    \frac{1}{m_{ee}} \mid F(m_{ee}^2) \mid^2 (1 - \frac{m_{ee}^2}{M^2})^3 \; ,
     \nonumber
    \label{eq:Wa}
  \end{eqnarray}
where the form factor, $F$, is unity for real photons.
Note that the phase space for Dalitz decays is limited by the
mass of the parent meson ($m_{ee} < M_{\pi^0,\eta,\omega}$),
while for direct photons it is not ($m_{ee}\sim p_T$).
Therefore, the measurement becomes relatively clean for
$p_T > 1$\,GeV/c, which is still in the low-$p_T$ realm where
the ``traditional'' calorimeter measurement has serious
difficulties.  The method is illustrated
on Fig.~\ref{fig:akiba}, described in detail in~\cite{bathe2005} 
and the resulting photon excess
ratios are shown as solid circles on Fig.~\ref{fig:photon_excess_AuAu}. 
The systematic errors are
much smaller for the internal conversion measurement (solid circles)
than for the traditional calorimeter measurement (open circles)
but the statistical errors become large for $p_T>4$\,GeV/c.
Tantalizingly, the low $p_T$ region is consistent with a 10\% photon 
excess over the hadronic background (and with the traditional
measurement), but before drawing conclusions about a thermalized
source the same measurement has to be repeated in p+p to establish 
the baseline.  Also, since internal conversion is a very small
probability process the method would greatly benefit from at least 
an order of magnitude increase of the available dataset 
(Sec.~\ref{sec_outlook}).

\begin{figure}
\begin{center}
\includegraphics*[width=8cm]{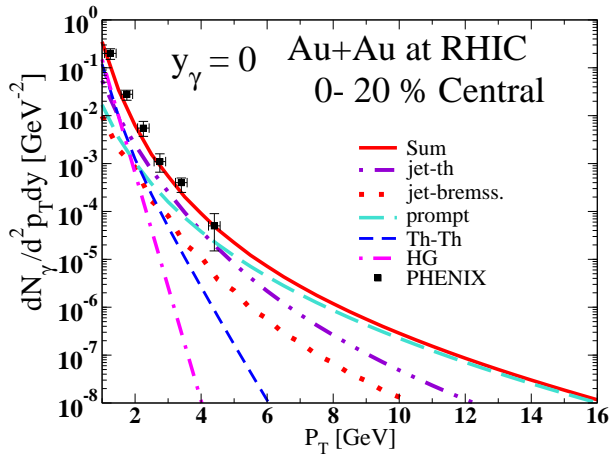}
\end{center}
\caption{Predictions for direct photon spectra combining
pQCD initial, jet-plasma interactions~\cite{turbide2005}
as well as thermal radiation~\cite{turbide2004}, compared
to preliminary PHENIX data.  The lines
labeled "Th-Th" and "HG" correspond to the usual thermal radiation
from QGP and hadron gas, respectively.
     }
\label{fig:phot-phenix-theo}
\end{figure}

Using the inclusive photon yields from the calorimeter 
measurement~\cite{ppg042} the excess ratio on 
Fig.~\ref{fig:photon_excess_AuAu} can be translated into a direct
photon yield and compared to theory as shown on 
Fig.~\ref{fig:phot-phenix-theo}.  Partial contributions are calculated
in~\cite{turbide2004,turbide2005} and the solid curve is the sum of
all subprocesses; it describes the data very well.  Remarkably, this
calculation uses an expanding fireball with $\tau_0=0.33$fm/c and
$T_i=370$MeV, double the formation time and only 2/3 of the
temperature conjectured in~\cite{denterria} and applied on
Fig.~\ref{fig:cdoubleratio_comp}.  In fact, five different models with
thermalization times in the range 0.15-0.5fm/c and initial temperatures
between 300MeV and 660MeV describe the data within a factor of 2.
But even with these uncertainties a very important lesson has already
been learned: while pre-RHIC expectations put the thermalization time
to $\sim1$fm/c, in turned out to be much shorter and the need to find
the mechanism of this unexpectedly rapid thermalization triggered a
large amount of theoretical work recently.

\section{Azimuthal asymmetries in direct photon production}
\label{sec_photonv2}

\begin{figure}
\begin{center}
\includegraphics*[width=8cm]{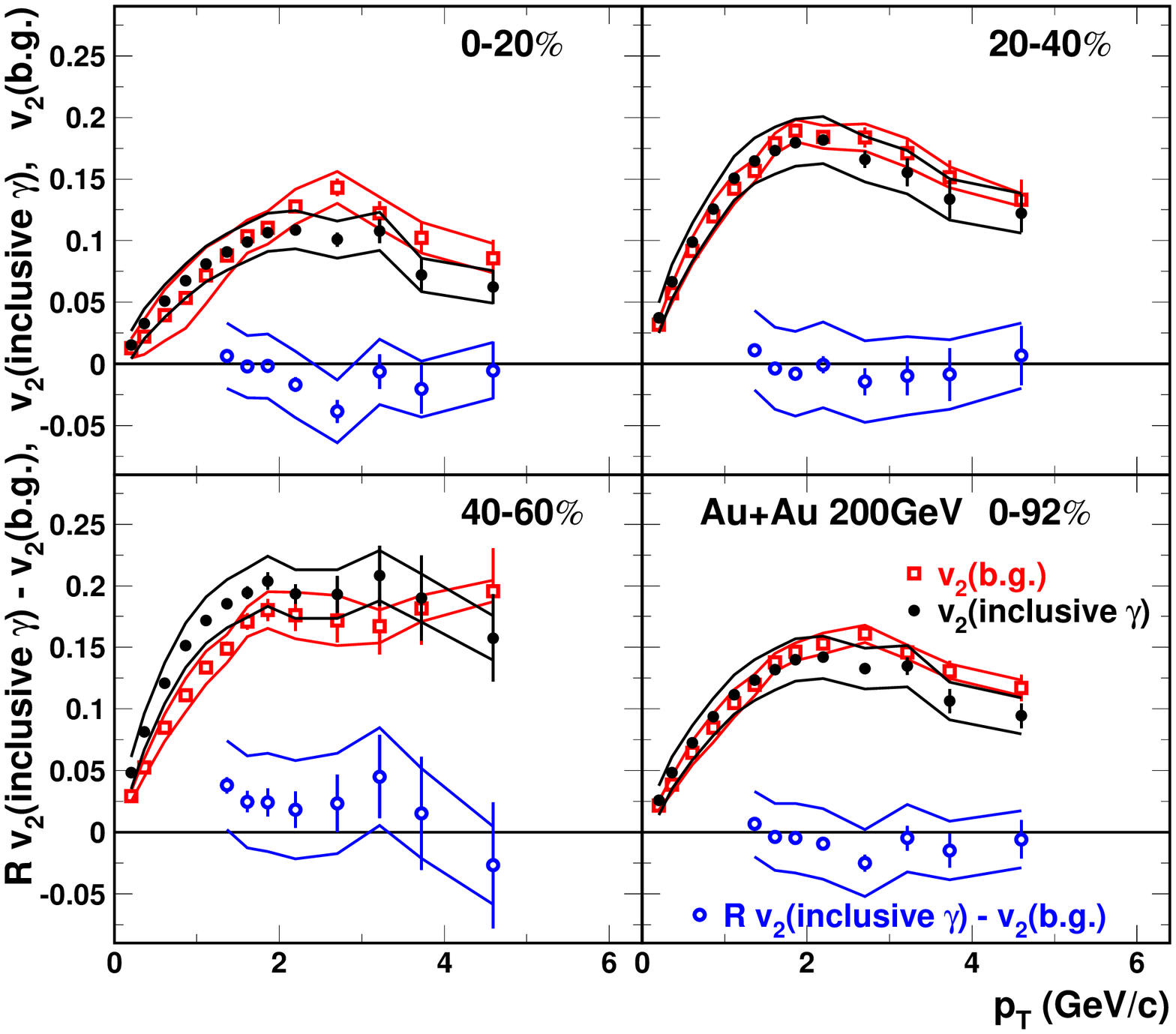}
\end{center}
\caption{Nuclear modification factor $R_{AA}$ in central Au+Au
  collisions ($\sqrt{s{NN}}=200$GeV) for $\pi^0$, $\eta$ and direct
  photons, compared to a GLV calculation with $dN^g/dy=1200$ gluon
  density.  The mesons - independent of their mass - are suppressed by
  a factor of 5 with respect to their yield in p+p scaled with the
  nuclear thickness $T_{AB}$, whereas the direct photons are not
  suppressed.   }
\label{fig:ppg046_fig2}
\end{figure}

We have seen in Fig.~\ref{fig:photon_excess_AuAu} 
that the {\it overall} yields of
high-$p_T$ photons (integrated over azimuth) scale with $T_{AA}$
and are well described by pQCD
(within current experimental and theoretical uncertainties).
However, this global agreement may mask more subtle effects
and it is even possible that the agreement is only accidental,
due to cancellations of processes that enhance and others that
quench the photon yield.  An important step toward clarification is to
study azimuthal asymmetries of photon distributions,
specifically their elliptic flow ($v_2$).
If (and since) the photons from the
initial scattering do not interact with the medium, their $v_2$
is expected to be zero. However,
initial hard scattering is far from being the only source of photons
in Au+Au collisions. Photons may also originate from jet partons
scattering off thermal partons (jet-thermal interactions)
or from Brehmsstrahlung off a quark.  These photons are expected
to exhibit a negative
$v_2$~\cite{turbide2006,chatterjee2005}, since more
material is traversed out-of-plane (which is the major axis in
coordinate space), with a strong $p_T$-dependence.
On the other hand, photons from thermal radiation should reflect the
dynamical evolution of the hot and dense matter thus carrying
a positive $v_2$.

A first measurement of photon elliptic flow is shown in
Fig.~\ref{fig:ppg046_fig2} taken from Ref.~\cite{ppg046}.  The
measurement is quite delicate due to the large background from
$\pi^0$ decay-photons that inherit their parent's $v_2$.
The measured $v_2$ of inclusive photons is consistent with
$v_2$ of photons from hadronic decays, i.e., a zero net direct
photon flow - but the error bars are appreciable and the
direct-to-inclusive photon ratio is very small at low $p_T$.
The quality of the data currently available is not sufficient to
prove or disprove theoretical
predictions~\cite{turbide2006,chatterjee2005}, not even the sign
of the net flow.  Much higher statistics can help remedy the situation,
at least at higher $p_T$: although the net direct photon flow is
predicted to decrease, the statistical errors will also become
smaller and, equally important, the direct-to-inclusive photon ratio
increases dramatically.  But even at high $p_T$ the net flow
will be a competition between processes with $v_2>0$ and $v_2<0$.
New analysis techniques may be able to disentangle (at least statistically)
isolated and non-isolated direct photons in heavy-ion collisions.
Jet-photon conversions produce mostly isolated
photons~\cite{turbide2006} with $v_2<0$ and the magnitude of this
flow depends strongly both on $p_T$ and the energy-loss mechanism
of jets in heavy-ion collisions. Therefore, a measurement of
the $v_2$ for isolated photons may give an independent constraint on
energy-loss models.

\section{Outlook}
\label{sec_outlook}

\begin{figure}
\begin{center}
\includegraphics*[width=8cm]{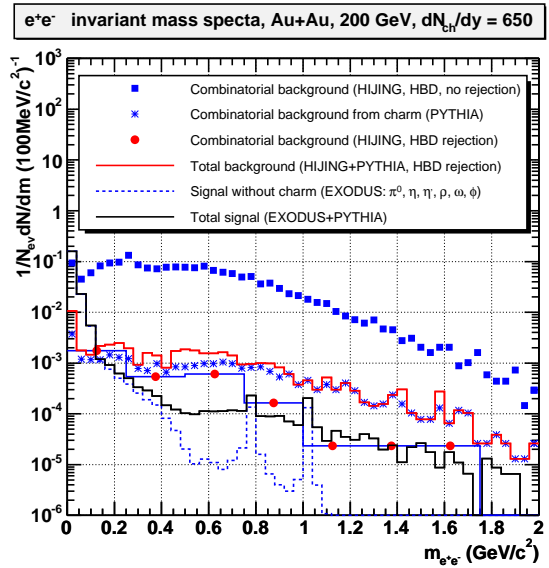}
\end{center}
\caption{
Combinatorial background for low-mass electron pairs compared to the
total signal from vector mesons and charm with and without the HBD.
Blue squares: total $e^+e^-$ combinatorial (HIJING), no rejection
from the HBD.  Red circles: combinatorial background after rejection
with HBD.  Blue stars: combinatorial background from charm alone
(PYTHIA), not rejected by the HBD, measured separately by the Silicon
Vertex Detector (SVTX).  Solid red line: total combinatorial background
after HBD rejection (HIJING+PYTHIA).  Dashed blue line: pure signal
from direct vector-meson and Dalitz decays after freezeout (``coktail"),
without contribution from charm (EXODUS).
Solid black line: total signal from vector mesons, Dalitz decays and
charm (EXODUS+PYTHIA).
     }
\label{fig:hbd_PHENIX}
\end{figure}

One of the fundamental questions in the pQCD phase transition is
whether chiral symmetry is (partially?) restored and in this context
to measure the in-medium modification of the light vector meson 
($\rho,\omega,\phi$) spectral functions.  In PHENIX this can be done
in the $e^+e^-$ channel, and ``proof of principle'' results have
been shown earlier~\cite{toia2005}.  However, suffering from an
$\sim 1/500$ signal/background ratio - combinatorics from $\pi^0$
Dalitz-decays and photon conversions - they are not precise enough to
make a meaningful statement on broadening or dropping of vector meson
masses.  A hadron-blind detector (HBD), already foreseen in the
original PHENIX design and actually installed September 2006 for Run-7
will remedy this situation, as illustrated by the simulations shown on
Fig.~\ref{fig:hbd_PHENIX}.  This novel detector, described
in~\cite{frankel} is a windowless Cherenkov detector operated with
pure CF$_4$ in a proximity focus configuration with a CsI photocathode
and a triple GEM detector with pad readout, and improves the signal to
background ratio to $\sim 1/10$.

The remaining background from
open charm will be measured in PHENIX separately with a Silicon Vertex
Detector (SVTX), which will measure the heavy-flavor
displaced vertex with a 40$\mu$m resolution of the distance of closest
approach, the specification driven by $c\tau$ of 123 and 462\,$\mu$m
for $D^0$ and $B^0$ decays, respectively.
The SVTX will have a central barrel and two endcap detectors,
thus covering both central and forward rapidities, providing
inner tracking with full azimuthal coverage and up to $|\eta|<2.4$.
This, in particular, will enable the
measurement of correlated $e\mu$ invariant-mass spectra and thus
provide for a stand-alone determination of the correlated open-charm
(and -bottom) component in the dilepton spectra.

Photons, $\pi^0$ and $\eta$ mesons can be measured well at
mid-rapidity in PHENIX, but the pseudorapidity coverage of the electromagnetic
calorimeter (along with the current ``central arm'') is limited
to $|\eta|<0.35$.  This makes full jet reconstruction very difficult.
Also, several measurements made at large rapities in d+Au
collisions suggest that in the low Bjorken-$x$ domain
gluons might be saturated and the CGC-model properly describes
the results, including hadron suppression at large rapidities
in d+Au (as opposed to no suppression as observed at $y=0$).
If the suppression at large $y$ is indeed a consequence of gluon
saturation (initial state), photons should also be suppressed
there - an important test feasible only
with a calorimeter at large
rapidities.  The limited acceptance of the central arm
also makes the crucial $\gamma$-jet measurements very difficult.
To remedy the situation PHENIX already added a high resolution
Muon Piston Calorimeter (MPC) to $3 < \eta < 4$  
and proposed to add a calorimeter
replacing the current copper nosecones of the magnet and covering
$1 < |\eta| < 3$.  
The longitudinally segmented Nose-Cone Calorimeter (NCC) will allow for
distinction between electromagnetic and hadronic showers.
Jet physics and energy-loss
studies using both photon-tagged jets and leading $\pi^0$'s will be
possible with the NCC, far away from central rapidity.  It will also
be very useful in studying heavy quarkonia enabling for instance the
measurement of $\chi_c \rightarrow \gamma J/\psi$ and possibly the
$\chi_b \rightarrow \gamma \Upsilon$ states in conjunction with the
existing muon spectrometer.  It 
is also an important addition to the spin program in measuring the
(polarized) gluon structure functions at low $x$.

\section{Summary}
\label{sec_summary}

In this paper we reported on recent measurements of direct photons by
the PHENIX experiment at RHIC at central rapidity.  Invariant yields
in p+p collisions are well described by NLO pQCD calculations, as are
the d+Au results if the calculations are scaled with the nuclear
thickness function.  In Au+Au the high transverse momentum region is
well reproduced by (the $T_{AA}$-scaled) NLO pQCD; at medium
transverse momenta additional, medium-induced production mechanisms
are possible although so far they are neither confirmed nor ruled out 
by measurements of
the azimuthal anisotropy of photons.  At low $p_T$ the Au+Au data are
consistent with a 10\% direct photons excess whose origin will be
constrained by a similar measurement in p+p.  The hadron-blind
detector currently installed in PHENIX will give access to high
quality data on low mass vector mesons. Current and mid-term upgrades 
will extend the photon and jet measurements to $1 < |\eta| < 4$.

% \appendix

% \section{}
% \label{}

\end{document}